\documentclass[english,aps,prc,showpacs,twocolumn,letter]{revtex4}

\usepackage[T1]{fontenc}
\usepackage[latin9]{inputenc}
\usepackage{graphicx}
\usepackage{esint}

\makeatletter
\@ifundefined{textcolor}{}
{%
 \definecolor{BLACK}{gray}{0}
 \definecolor{WHITE}{gray}{1}
 \definecolor{RED}{rgb}{1,0,0}
 \definecolor{GREEN}{rgb}{0,1,0}
 \definecolor{BLUE}{rgb}{0,0,1}
 \definecolor{CYAN}{cmyk}{1,0,0,0}
 \definecolor{MAGENTA}{cmyk}{0,1,0,0}
 \definecolor{YELLOW}{cmyk}{0,0,1,0}
}

\usepackage{babel}

\makeatother

\usepackage{babel}
\begin{document}

\title{Decorrelation of anisotropic flows along the longitudinal direction}

\author{Long-Gang Pang$^1$, Hannah Petersen$^{1,2,3}$, Guang-You Qin$^{4}$, Victor Roy$^2$ and Xin-Nian Wang$^{4,5}$, }

\address{$^1$Frankfurt Institute for Advanced Studies, Ruth-Moufang-Strasse 1, 60438 Frankfurt am Main, Germany}
\address{$^2$Institute for Theoretical Physics, Goethe University, Max-von-Laue-Strasse 1, 60438 Frankfurt am Main, Germany}
\address{$^3$GSI Helmholtzzentrum f\"ur Schwerionenforschung, Planckstr. 1, 64291 Darmstadt, Germany}
\address{$^4$Key Laboratory of Quark \& Lepton Physics (MOE) and Institute of Particle Physics, Central China Normal University, Wuhan 430079, China }
\address{$^5$Nuclear Science Division MS70R0319, Lawrence Berkeley National Laboratory, Berkeley, CA 94720}

\begin{abstract}
The initial energy density distribution and fluctuation in the transverse direction lead to anisotropic flows
of final hadrons through collective expansion in high-energy heavy-ion collisions. 
Fluctuations along the longitudinal direction, on the other hand, can result in
decorrelation of anisotropic flows in different regions of pseudo rapidity ($\eta$).
Decorrelation of the $2$nd and $3$rd order anisotropic flows with different $\eta$ gaps 
for final charged hadrons in high-energy heavy-ion collisions
is studied in an event-by-event (3+1)D ideal hydrodynamic model with fully fluctuating initial 
conditions from A Multi-Phase Transport (AMPT) model. 
The decorrelation of  anisotropic flows of final hadrons with large $\eta$ gaps 
are found to originate from the spatial decorrelation along the longitudinal direction
in the AMPT initial conditions through hydrodynamic evolution. The decorrelation is found to
consist of both a linear twist and random fluctuation of the event-plane angles.
The agreement between our results and recent CMS data in most
centralities suggests that the string-like mechanism of initial parton production in AMPT
model captures the initial longitudinal fluctuation that is responsible for the measured 
decorrelation of anisotropic flows in Pb+Pb collisions at LHC. Our predictions for Au+Au collisions 
at the highest RHIC energy show stronger longitudinal decorrelation, indicating larger longitudinal 
fluctuations at lower beam energies. Our study also calls into question some of the current experimental
methods for measuring anisotropic flows and extraction of transport coefficients through
comparisons to hydrodynamic simulations that do not include longitudinal fluctuations. 
\end{abstract}



\maketitle

\section{Introduction}

Anisotropic flows or Fourier coefficients  $\vec{V}_{n}=v_{n}\exp(in\Psi_{n})$ of 
the distribution of final charged hadrons in azimuthal angle 
are important observables in high-energy heavy-ion collisions.
They provide critical information about the initial state and evolution of the strongly coupled Quark
Gluon Plasma (QGP). They have been used to extract the ratio of shear viscosity over entropy
density $\eta_{v}/s$ of the QGP through comparisons between event-by-event
viscous hydrodynamic calculations and experimental measurements at
the Relativistic Heavy-ion Collider (RHIC)  and the Large Hadron Collider (LHC)  \cite{Romatschke:2007mq,Luzum:2008cw,Song:2007fn,Song:2007ux,Dusling:2007gi,Molnar:2008xj,Bozek:2009dw,Chaudhuri:2009hj,Schenke:2010rr,Schenke:2011bn}.
However, it is well known that $v_{n}$'s from hydrodynamic simulations are very sensitive to initial
state fluctuations \cite{Hirano:2009ah,Alver:2010gr}. For example,
values of $\eta_{v}/s$ extracted from fitting experimental data on $v_2$ with
viscous hydrodynamic simulations differ between the Monte Carlo Glauber (MC-Glauber) and Monte
Carlo Color Glass Condensate (MC-CGC) models for initial state conditions, while all other 
parameters in the viscous hydrodynamic model are kept fixed \cite{Alver:2010dn,Adare:2011tg,Retinskaya:2013gca}.
In addition, many models of initial conditions for viscous hydrodynamic simulations fail to describe the second ($v_{2})$
and the third ($v_{3}$) order harmonic flow coefficients simultaneously in ultra central collisions \cite{Shen:2015qta}. 
The transverse momentum and pseudo rapidity dependent event plane angles $\Psi_{n}(p_{T},\eta)$
in $3+1$D hydrodynamic simulations including full fluctuations in both transverse and longitudinal
directions have not been fully investigated yet.

The initial energy density distribution and fluctuations of the QGP in the
transverse plane have been studied in detail in event-by-event hydrodynamics
with initial conditions given by MC-Glauber, MC-CGC \cite{Broniowski:2007ft, Hirano:2009ah},
UrQMD \cite{Petersen:2008dd}, EPOS \cite{Werner:2010aa}, AMPT
\cite{Pang:2012he} and IP-Glasma \cite{Gale:2012rq}. Fluctuations in the transverse plane not
only give rise to odd flow harmonics but also significant even and odd $v_n$ in ultra central 
collisions \cite{ Ma:2010dv}. They also result in $p_{T}$
dependent event planes, which break down the flow factorization 
$v_{n,n}(p_{T1},p_{T2})=v_{n}(p_{T1})v_{n}(p_{T2})$
\cite{Aamodt:2011by,Heinz:2013bua,Gardim:2012im,Qiu:2012uy,Qiu:2013wca},
where $v_{n,n}$ is obtained from the long-range two-particle correlation function.
The decorrelation between event planes for particles with fixed transverse
momentum $p_{T1}$ and the event plane determined by particles in
the full $p_{T}$ range may even give rise to negative $v_{n}$ at
$p_{T1}$ for ultra central collisions.

Studies of fluctuations along the longitudinal direction and their effects on anisotropic flows of 
final charged hadrons have only recently been started. Energy density fluctuations
in spatial rapidity were first studied in a Boltzmann parton and hadron transport + hydrodynamic
hybrid approach \cite{Petersen:2011fp}. Later it was observed that
the elliptic flow $v_2$ is noticeably suppressed in the presence of longitudinal
fluctuations in event-by-event (3+1)D ideal hydrodynamics with AMPT initial
conditions \cite{Pang:2012he}. This was later confirmed by another
independent study within the AMPT model \cite{Xiao:2012uw} in which the pseudo rapidity
window for event plane determination has been varied. A comparison between results from (3+1)D ideal hydrodynamics and
direct AMPT model simulations shows that both models convert the fluctuation in the initial energy
momentum tensor in coordinate space to a decorrelation
of anisotropic flows for final charged hadrons in momentum space \cite{Pang:2014pxa}.
These early studies indicate that the longitudinal structure of the initial states of QGP and their evolution 
play an important role in extracting the transport coefficients from experimental data and may shed light on the $v_{n}$
puzzle in ultra central collisions.

Variations of event plane angles $\Psi_{n}(\eta)$ in the longitudinal direction can consist of a linear
twist and random fluctuations along the $\eta$ direction. The linear twist is a continuous rotation  
of event plane angles from project to target beam direction in heavy-ion collisions.
The twist in gluon number density distributions was first suggested in the CGC model
\cite{Adil:2005bb,Adil:2005qn}, considering the trapezoidal distribution
of gluon number density $\rho_{g}(\eta,x_{\perp})$ associated with
asymmetric forward and backward participants in projectile and
target nuclei in non-central heavy-ion collisions. The twist of event plane angles was 
also suggested by Bozek \cite{Bozek:2010vz} in a wounded-nucleon model, where
fireballs in relativistic heavy-ion collisions may be torqued with
the assumption that the forward (backward) wounded nucleons
emit particles preferably in the forward (backward) direction. The
twist in the energy density distribution along the longitudinal direction
in the initial state will lead to a torqued collective flow and torqued
momentum distributions in different rapidity windows.  On top of the
twist, there is also a random fluctuation of event plane angles $\Psi_{n}(\eta)$
due to the finite number of particles in a given window of pseudo-rapidity. 
Both the twist and random fluctuations in the initial energy density distribution lead to 
variations of event plane angles $\Psi_{n}(\eta)$ along the longitudinal direction and 
decorrelation of anisotropic flows of final hadrons with large pseudo-rapidity gaps.
It is important to separate effects of these two mechanisms for 
decorrelation of anisotropic flows along the longitudinal direction.

%However, the random fluctuations break the Bjorken
%scaling in each single event and the plateau structure in $(p_{T},\phi)$
%integrated $dN/d\eta$ at middle rapidity is the result of the event
%average. While the twist of event plane angles only break the Bjorken
%scaling locally and keep Bjorken scaling for the $(p_{T},\phi)$ integrated
%$dN/d\eta$ in each single event. This difference can be used as one
%necessary condition to separate random fluctuations from twist effect
%experimentally. (There is another random fluctuation picture from
%CGC, granularity difference in different rapidity windows).

%Many ways are suggested to measure the fluctuation or twist along
%the pseudo rapidity direction. For hard probes, 3D jet tomography
%is suggested to measure the twist of strongly coupled QCD matter predicted
%from the CGC model \cite{Adil:2005bb,Adil:2005qn}. 

Many techniques have been proposed to study the longitudinal structure of
final hadron production in heavy-ion collisions and the underlying mechanisms.
For example,  three particle correlations were suggested to measure the twist 
effect \cite{Borghini:2002hm} in heavy-ion collisions at RHIC. 
One can also characterize the longitudinal fluctuation in terms of
coefficients in the Legendre polynomial expansion of two-particle correlations in 
pseudo-rapidity \cite{ATLAS:2015kla,Monnai:2015sca,Bozek:2015tca}.
The most intuitive method is to measure the forward-backward event plane
angles \cite{Bozek:2010vz,Xiao:2012uw, Huo:2013qma} or anisotropic flow differences
\cite{Pang:2014pxa} with varying pseudo rapidity gaps $\Delta\eta$.
These methods are used within the torqued fireball model \cite{Bozek:2010vz},
(3+1)D hydrodynamics model \cite{Pang:2014pxa,Bozek:2010vz} and the
AMPT model \cite{Jia:2014ysa,Pang:2014pxa,Xiao:2012uw} to study
the decorrelation of event plane angles or anisotropic flows along
the pseudo rapidity direction.  Jia et el. \cite{Huo:2013qma,Jia:2014ysa}
also proposed an ``event-shape twist'' technique to study the event plane decorrelation due to
the twist in initial energy density distributions by selecting events with
big forward-backward (FB) event plane angle differences. By selecting
events with vanishingly small FB event plane angle differences,
one can then eliminate the twist effect and the measured decorrelation
of anisotropic flows with finite pseudo rapidity gaps should be caused
only by random fluctuations of event plane angles \cite{Pang:2014pxa}.

The most recent CMS measurements \cite{Khachatryan:2015oea} use a different
definition of correlations of anisotropic flows which intends to remove contributions from
short range correlations that can arise from the longitudinal expansion of hot spots along $\eta$,
non-flow correlations from jet fragments, Bose-Einstein correlation and resonance decays. 
In this paper, we carry out a study of such correlations within an event-by-event (3+1)D ideal 
hydrodynamic model with AMPT initial conditions and compare with CMS experimental data. 
We will investigate the decorrelation of 2nd and 3rd order anisotropic flows along the pseudo-rapidity
direction. The rest of this article is organized as follows. In Sec.
\ref{sec:method} we introduce the definition of the longitudinal
decorrelation, the (3+1)D ideal hydrodynamic model and the fluctuating
initial conditions employed in our study. Results for Pb+Pb collisions
at LHC are compared with CMS experimental data and predictions for Au+Au collisions at RHIC 
are presented  in Sec. \ref{sec:results}.  In Sec. \ref{sec:initial_state_decorr}
we investigate fluctuations and the decorrelation in the initial state
in coordinate space. The twist of event plane angles and random fluctuations
are studied in Sec. \ref{sec:twist_or_fluctuation} for all centralities
to illustrate the non-linear decorrelation in most central collisions.
A summary and discussions on effects of longitudinal fluctuations on anisotropic flow and di-hadron
correlations are given in Sec.\ref{sec:discussion}.

\section{Method and Model \label{sec:method}}

\subsection{Decorrelation of anisotropic flow in pseudo rapidity}

We start with the definition of the correlation observable \cite{Khachatryan:2015oea} proposed
by CMS. The pesudo-rapidity coverages of $\eta^{a}\in(-2.4,\ 2.4)$ for Pb+Pb collisions 
at $\sqrt{s_{NN}}=2.76$ TeV and $\eta^{a}\in(-1.5,\ 1.5)$ for Au+Au collisions at $\sqrt{s_{NN}}=200$ GeV
are divided into $16$ and $10$ rapidity bins, respectively, with equal bin size $\Delta\eta=0.3$.
Particles from reference pseudo-rapidity windows $\eta_{b}\in(3.0,\ 5.0)$
and $\eta_{b}\in(2.5,\ 4.0)$ are used in Pb+Pb and Au+Au collisions,
respectively, to be correlated with particles at $\eta^{a}$ which
is denoted as $V_{n\Delta}(\eta^{a},\eta^{b})$ to remove short
range correlations in the denominator. The ratio between $V_{n\Delta}(-\eta^{a},\eta^{b})$
and $V_{n\Delta}(\eta^{a},\eta^{b})$ (for $\eta_a>0$) serves as a measure of the decorrelation
of anisotropic flows as a function of pseudo rapidity gap $\Delta\eta=2\eta^{a}$,
\begin{eqnarray}
& & r_{n}(\eta_{a},\eta_{b}) = V_{n\Delta}(-\eta_{a},\eta_{b}) / V_{n\Delta}(\eta_{a},\eta_{b}) \nonumber \\
                                     &=& \frac{\left\langle v_{n}(-\eta_{a})v_{n}(\eta_{b})\cos\left[n\left(\Psi_{n}(-\eta_{a})-\Psi_{n}(\eta_{b})\right)\right]\right\rangle }
                                     {\left\langle v_{n}(\eta_{a})v_{n}(\eta_{b})\cos\left[n \left(\Psi_{n}(\eta_{a})-\Psi_{n}(\eta_{b})\right)\right]\right\rangle },
\end{eqnarray}
which is also called $\eta$-dependent factorization
ratio. The di-hadron correlation function factorizes along $\eta$
and $r_{n}(\eta_{a},\eta_{b})=1$ if there are no longitudinal fluctuations.
Otherwise, the correlation between $\eta_a$ and $\eta_b$ is stronger than that
between $-\eta_a$ and $\eta_b$, and the factorization in $\eta$ breaks down ($r_n(\eta_a, \eta_b)$ < 1).

A global twist of event planes or decreasing long range correlation with increasing
pseudo rapidity gap $\Delta\eta$ can lead to this kind of breaking of factorization.
Events with pure random fluctuations of $v_{n}(\eta)$ and $\Psi_{n}(\eta)$ do not contribute either to
numerator or the denominator of $r_{n}(\eta_{a},\eta_{b})$.

In practice we use the $\vec{Q}_{n}$ vector to quantify the $n$th
order anisotropic flow in a given pseudo-rapidity bin which is defined as,
\begin{eqnarray}
\vec{Q}_{n} & \equiv & Q_{n}e^{in\Phi_{n}}=\frac{1}{N}\sum_{j=1}^{N}e^{in\phi_{j}} \nonumber \\
 & = & \frac{\int\exp(in\phi)\frac{dN}{d\eta dp_{T}d\phi}dp_{T}d\phi}{\int\frac{dN}{d\eta dp_{T}d\phi}dp_{T}d\phi},
\end{eqnarray}
where $\phi_{j}=\arctan(p_{yj}/p_{xj})$ is the azimuthal angle of
the $j$th particle as determined by its transverse momentum. In
hydrodynamic simulations, smooth particle spectra $dN/d\eta p_{T}dp_{T}d\phi$ and phase space
integration over $p_{T}$ and azimuthal angle $\phi$ are used to
calculate the $\vec{Q}_{n}$ vector. In this case, the $\vec{Q}_{n}$ vector will be identical
to the flow vector $\vec{v}_{n}$ and the decorrelation of anisotropic
flows in $2$ different rapidity bins $-\eta_{a}$ and $\eta_{a}$
becomes,
\begin{equation}
r_{n}(\eta_{a},\eta_{b})=\frac{\left\langle \vec{Q}_{n}(-\eta_{a})\vec{Q}_{n}^{*}(\eta_{b})\right\rangle }{\left\langle \vec{Q}_{n}(\eta_{a})\vec{Q}_{n}^{*}(\eta_{b})\right\rangle }\label{eq:r_n_final}
\end{equation}
For collisions at both RHIC and LHC energies we follow the CMS analysis and use the transverse momentum cut $[0.3,\ 3.0)$
GeV/c for particles in pseudo-rapidity windows $\eta_{a}$ and $[0,\ \infty)$
for particles in $\eta_{b}$ .

\subsection{Event-by-event (3+1)D hydrodynamics}

A (3+1)D ideal hydrodynamical model \cite{Pang:2012he,Pang:2014ipa} is employed to study 
the decorrelation of anisotropic flows in different rapidity windows in Pb+Pb collisions at $\sqrt{s_{NN}}=2.76$ TeV 
and Au+Au collisions at $\sqrt{s_{NN}}=200$ GeV with fluctuating initial conditions from AMPT \cite{Lin:2004en}.
The initial state local energy momentum tensor is constructed from the 
coordinates $(t_{i},x_{i},y_{i},z_{i})$ and four-momenta
$(m_{Ti}\cosh Y_{i},p_{xi},p_{yi},m_{Ti}\sinh Y_{i})$ of partons when they
cross the $\tau_{0}=\sqrt{t^{2}-z^{2}}$ hyperbolic line. 
The initial energy momentum tensor is constructed in $(\tau,x,y,\eta_{s})$ coordinates
according to 
\begin{eqnarray}
&T^{\mu\nu}&(\tau_0, x, y, \eta_s) =  K\sum_{i}\frac{p_{i}^{\mu}p_{i}^{\nu}}{p_{i}^{0}} \frac{1}{\tau_0\sqrt{2\pi\sigma_{\eta_s}^2}} \frac{1}{2\pi\sigma_r^2} \nonumber \\
 &\times& \exp\left[-\frac{(x-x_{i})^2+(y-y_{i})^2}{2\sigma_r^2} - \frac{(\eta_{s}-\eta_{si})^2}{2\sigma_{\eta_s}^2}\right],
\label{eq:initial}
\end{eqnarray}
where,
\[
p_{i}^{\mu}=\left(m_{Ti}\cosh(Y_{i}-\eta_{s}),p_{xi},p_{yi},m_{Ti}\sinh(Y_{i}-\eta_{s})/\tau_{0}\right)
\]
are four-momenta of partons in $(\tau,x,y,\eta_{s})$ coordinates.
Local thermalization is assumed in coordinate space through a normalized
Gaussian distribution with widths $\sigma_{r}$ in transverse coordinates $r$ and $\sigma_{\eta_{s}}$
in spatial rapidity  $\eta_{s}$. The same smearing widths $\sigma_{r}=0.6$ fm and $\sigma_{\eta_{s}}=0.6$
as in previous studies \cite{Pang:2012he} are used in this calculation, and different values
of $\sigma_{\eta_{s}}=0.4,\ 0.6,\ 0.8$ are employed to study the
sensitivity of the factorization ratio $r_{n}(\Delta\eta)$ to short range correlations
from smearing. The partial chemical equilibrium equation of state
(EOS) s95p-PCE-v0 from lattice QCD calculations \cite{Huovinen:2009yb}
is used in the hydrodynamic model. This EOS is known to produce softer
transverse momentum spectra and bigger anisotropic flows than the chemical
equilibrium EOS \cite{Pang:2012he}. A parameter $K$ in Eq.~(\ref{eq:initial}) is used to fit the final charged hadron
multiplicity at mid-rapidity for most central collisions. With the s95p-PCE-v0  EOS
, one finds $K=1.5$ for LHC energy and $K=1.4$ for RHIC
energy. The same values are used for all other centralities. The initial thermalization
time $\tau_{0}$ is unchanged from previous simulations \cite{Pang:2012he} where $\tau_{0}=0.2$
fm for LHC and $\tau_{0}=0.4$ fm for RHIC energy.

AMPT uses HIJING \cite{Wang:1991hta} to generate initial partons
from hard and semi-hard scatterings and excited strings from soft
interactions. The number of mini-jet partons per binary nucleon-nucleon
collision in hard and semi-hard scatterings follow a Poisson distribution
with the mean value given by the jet cross section. The number of
excited strings is equal to the number of participant nucleons in
each event.  The AMPT model uses a string-melting mechanism to convert strings into
partons that will follow a parton cascade and eventually hadronize according to a
parton recombination model. We run AMPT in Cartesian coordinates to the end and
extract the initial condition at a given initial invariant time $\tau_0$ for our hydrodynamic simulations.
The Monte Carlo Glauber model is used in HIJING to determine
the number of binary collisions and the number of participants. Besides
 random fluctuations from mini-jet partons, the parton density fluctuates
along longitudinal direction according to the length of strings. There are basically
three types of strings,
\begin{itemize}
\item strings associated with each wounded nucleon (between a valence quark and a diquark),
\item single strings between $q-\bar{q}$ pairs from quark annihilation and gluon fusion processes,
\item strings between one hard parton from parton scatterings and valence
quark or di-quark in wounded nucleons.
\end{itemize}
Fluid expansion of the tube-like energy density from these strings
results in long range correlations in $\eta$ (ridge) in the final state.
Fluctuations of string lengths together with the asymmetric
distribution between forward-backward participants provide
large fluctuations along the longitudinal direction. Length fluctuations
of color flux tubes are also assumed by Bozek in the torqued fireball
model in order to explain the breakdown of factorization along $\eta$
in p+Pb collisions \cite{Bozek:2015bna}. 

For careful comparisons with experimental data involving event-by-event
fluctuations, centrality selections have to be chosen consistently with similar event classes.  
In the CMS experiment \cite{Khachatryan:2015oea}, centrality classes in 
Pb+Pb collisions at  $\sqrt{s_{NN}}=2.76$ TeV/n are determined by the total energy deposited in 
the hadronic forward (HF) calorimeters, and the results are equivalent to those obtained by total 
number of final charged hadrons. In our calculations in this paper, we determine
centrality classes by the number of initial partons, which is proportional to the number of final 
charged hadrons after hydrodynamic simulations.

\begin{figure}[!tph]
\includegraphics[width=0.5\textwidth]{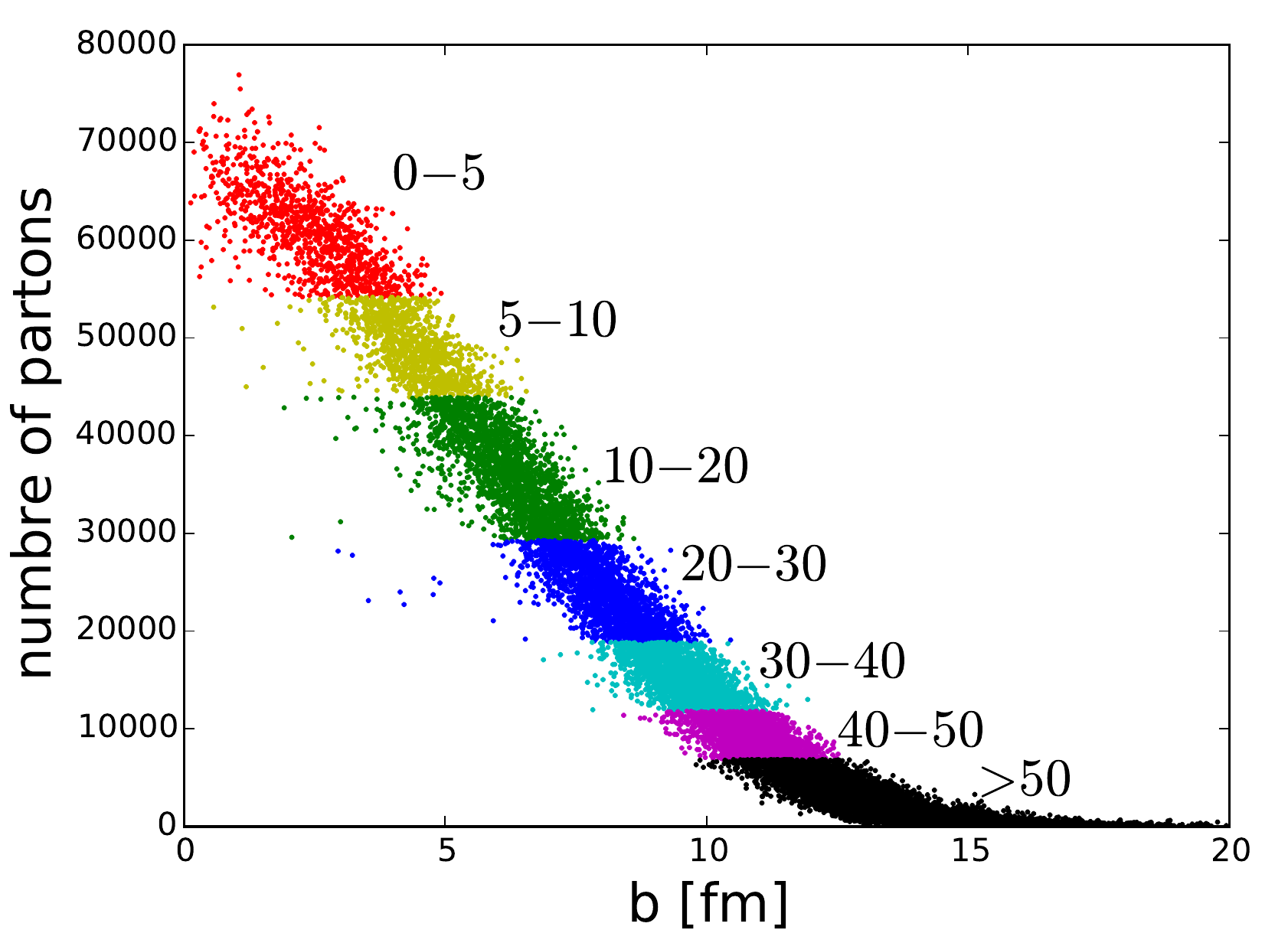} 
\protect\caption{(Color online) The centrality classes determined by number of initial
partons versus the impact parameter ranges. \label{fig:npartons_vs_b}}
\end{figure}

We use $100,000$ minimum bias AMPT events within the range $[r_{min},\ r_{max}]=[0,\ 20]$
fm of the impact parameter to provide initial conditions on a hyperbolic surface at an initial time $\tau=\tau_{0}$. 
Events are grouped into different centrality classes according to the percentage of events ordered by the total 
number of initial partons. For example,  events in the 0-5\% centrality are selected from the top 5\% of events
that have the highest number of initial partons, and so on. The scatter plot in Fig.~\ref{fig:npartons_vs_b}
shows different centrality classes determined from this method and their corresponding impact parameters.
Obviously, centrality classes determined by the number of initial partons are quite different
from those determined by ranges of  impact parameters. 

By selecting similar events for each centrality class as the CMS experiment at LHC, our current
simulations agree with experimental data better than the previous  results with centralities 
determined by ranges of impact parameters \cite{Pang:2014pxa}.

\begin{figure}[!tph]
\includegraphics[width=0.5\textwidth]{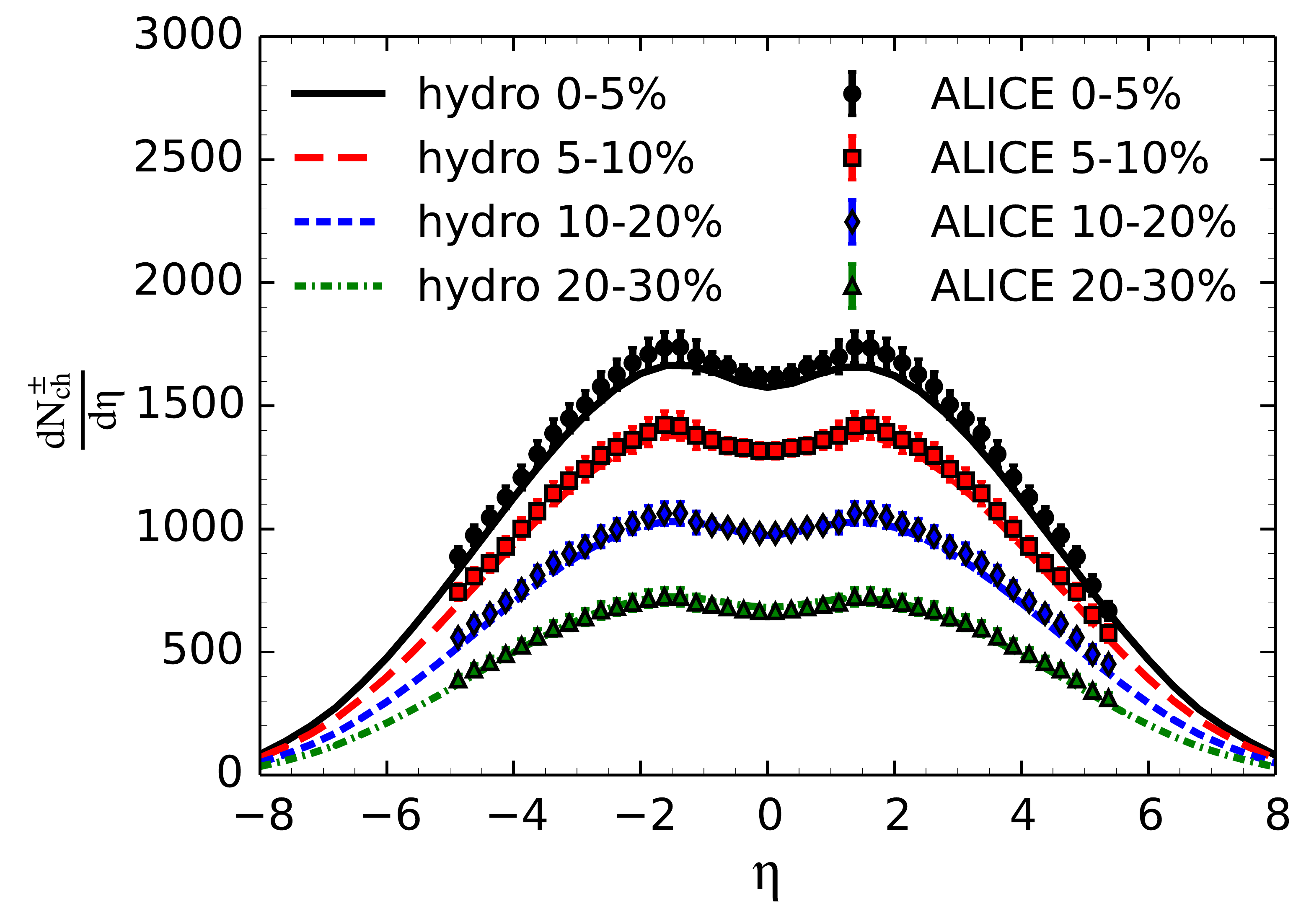}
\protect\caption{(Color online) Charged multiplicity distributions as a function of
pseudo rapidity for Pb+Pb $\sqrt{s_{NN}}=2.76$ TeV collisions from
event-by-event (3+1)D hydrodynamics compared with ALICE measurements at LHC.}
\label{fig-dndeta}
\end{figure}

We fix the parameters in our event-by-event hydrodynamic model by comparing to
experimental data on $dN_{ch}/d\eta$ for the 0-5\% central collisions .
The charged multiplicity as a function of pseudo rapidity $dN_{ch}/d\eta$ in other
centralities from our event-by-event hydrodynamic simulations agree with 
ALICE measurements \cite{Abbas:2013bpa} rather well as shown in Fig.~\ref{fig-dndeta}.
%Notice that if we adjust parameters so that $dN_{ch}/d\eta$ from event-by-event
%hydrodynamics for the most central collisions fits perfectly with experimental
%data, hydrodynamic calculations for semi-central and peripheral collisions
%will overestimated charged multiplicity a little bit. This mismatch
%is inherited from the AMPT initial conditions \cite{Abbas:2013bpa}. 

\begin{figure}[!tph]
\includegraphics[width=0.5\textwidth]{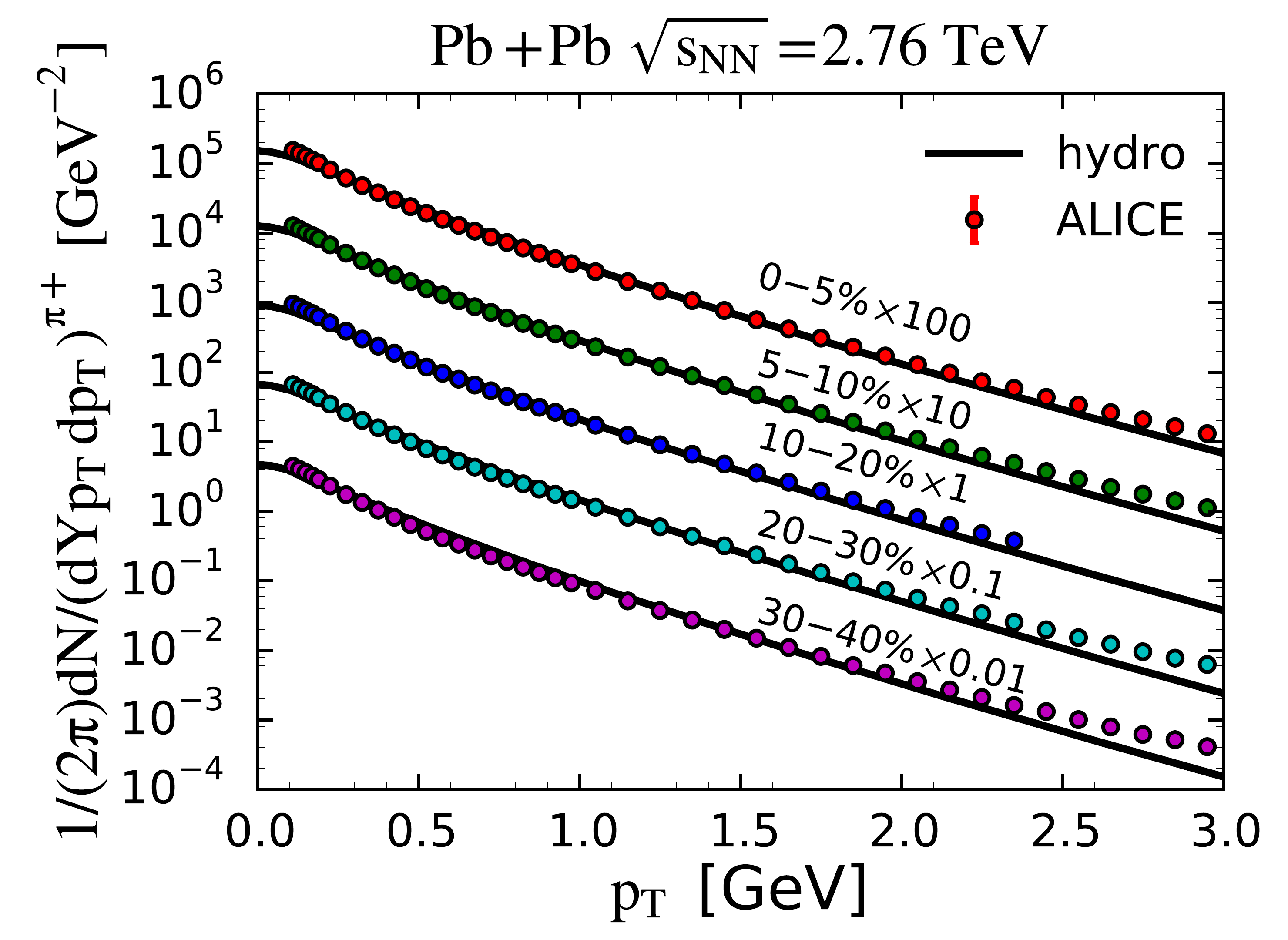}
\protect\caption{(Color online) Transverse momentum spectra of $\pi^+$ in Pb+Pb collisions
at $\sqrt{s_{NN}}=2.76$ TeV in different centralities from event-by-event (3+1)D
hydrodynamics (solid lines) compared with ALICE measurements at LHC (solid circles).
 \label{fig:ptspec_all_cents}}
\end{figure}

Shown in Fig.~\ref{fig:ptspec_all_cents} are transverse momentum spectra for $\pi^{+}$ in 
centrality classes $0-5\%$, $5-10\%$, $10-20\%$ , $20-30\%$ and $30-40\%$ from our 
event-by-event hydrodynamic calculations (solid lines)
as compared with ALICE measurements (solid circles).
Our calculations with (3+1)D ideal hydrodynamic model with partial chemical equilibrium
EOS agree very well with the experimental data especially in the
low $p_T$ region.  They, however, under-estimate the particle production at high $p_{T}$.
%which is expected to improve if finite viscosity is taken into account.
High $p_T$ spectra
are not expected to influence $p_T$-integrated anisotropic flows that are used to
calculate the factorization ratio $r_{n}(\Delta\eta)$.

%A good fit for $p_{T}$ spectra in the transverse momentum range $[0.3,3.0)$
%GeV/c provides the foundation for $r_{n}(\Delta\eta)$ calculations
%for particles in the pseudo rapidity range $[-\eta_{a},\eta_{a}]$.

\section{Results on factorization ratios \label{sec:results}}

The factorization ratios $r_{2}$ and $r_{3}$ as a function of $\eta_{a}$ in $6$ centralities from $0-5\%$ to $40-50\%$
collisions are shown in Figs.~\ref{fig:r2} and \ref{fig:r3}, for Pb+Pb collisions at $\sqrt{s_{NN}}=2.76$ TeV and 
Au+Au collisions at $\sqrt{s_{NN}}=200$ GeV from our event-by-event (3+1)D hydrodynamic simulations,
and compared with CMS measurements for Pb+Pb collisions at $\sqrt{s_{NN}}=2.76$
TeV \cite{Khachatryan:2015oea}. The factorization ratio for the second anisotropic flow $r_{2}(\eta_{a},\eta_{b})$
with the reference rapidity window $4.4<\eta_{b}<5.0$ from event-by-event hydrodynamic simulations 
show a rather nice agreement with CMS measurements for all centralities except $0-5\%$ central
collisions. The splitting of the factorization ratios  with $2$ different reference windows
$\eta_{b}$ from hydrodynamic calculations is tiny. This suggests that
the current definition of the factorization ratio is insensitive to short range correlations from
the hydrodynamic expansion of hot spots and resonance decay. Both
results fit better the CMS measurements with the reference rapidity
window $4.4<\eta_{b}<5.0$, where short range jet-like correlations
are strongly suppressed. For $0-5\%$ central collisions, the
decorrelation from event-by-event hydrodynamic simulations is more
linear than CMS measurements. Hydrodynamic predictions for RHIC energy have
quite similar centrality dependence, with much stronger decorrelation
along pseudo-rapidty than LHC energy. This is reasonable since
fluctuations at RHIC energy are much bigger than LHC energy. The prediction
is consistent with anisotropic flow measurements at RHIC and LHC,
where the variation of $v_{2}$ in $|\eta|<2.5$ is much bigger for
RHIC than LHC \cite{ATLAS:2011ah}.

\begin{widetext}

\begin{figure}[!tph]
\includegraphics[width=0.7\textwidth]{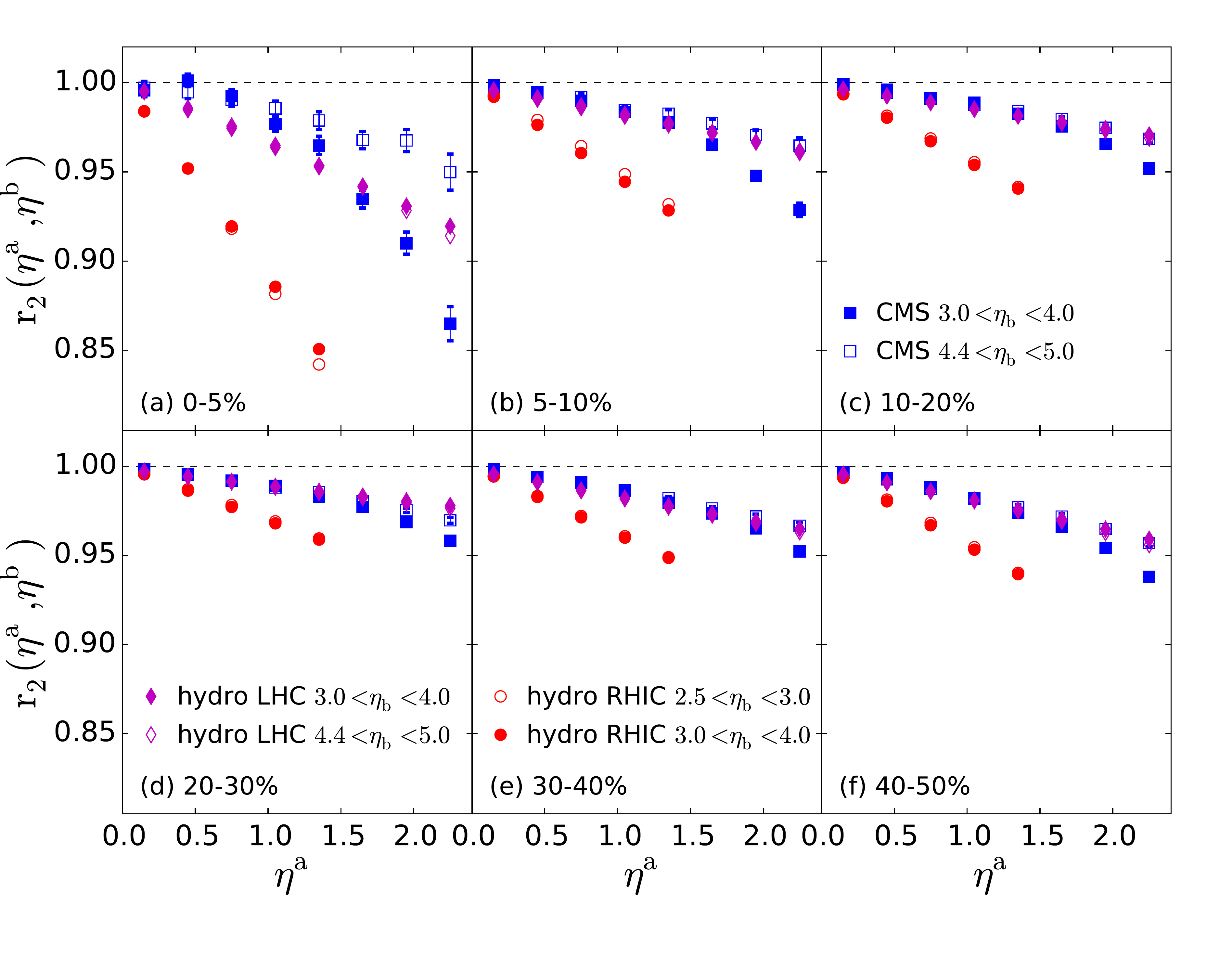} 
\protect\caption{(Color online) The factorization ratio $r_{2}$
as a function of $\eta^{a}$ for $3.0<\eta^{b}<4.0$ and $4.4<\eta^{b}<5.0$,
in Pb+Pb  collisions at $\sqrt{s_{NN}}=2.76$ TeV (open and solid diamonds),
and for $2.5<\eta^{b}<3.0$ and $3.0<\eta^{b}<4.0$ in Au+Au collisions at $\sqrt{s_{NN}}=200$
GeV  (open and solid circles) from event-by-event $3+1$D
ideal hydrodynamic simulations as compared with CMS experimental data
\cite{Khachatryan:2015oea} for Pb+Pb collisions at $\sqrt{s_{NN}}=2.76$ TeV/n
 (empty and solid squares) in $6$ different centralities.
\label{fig:r2}}
\end{figure}

\begin{figure}[!tph]
\includegraphics[width=0.7\textwidth]{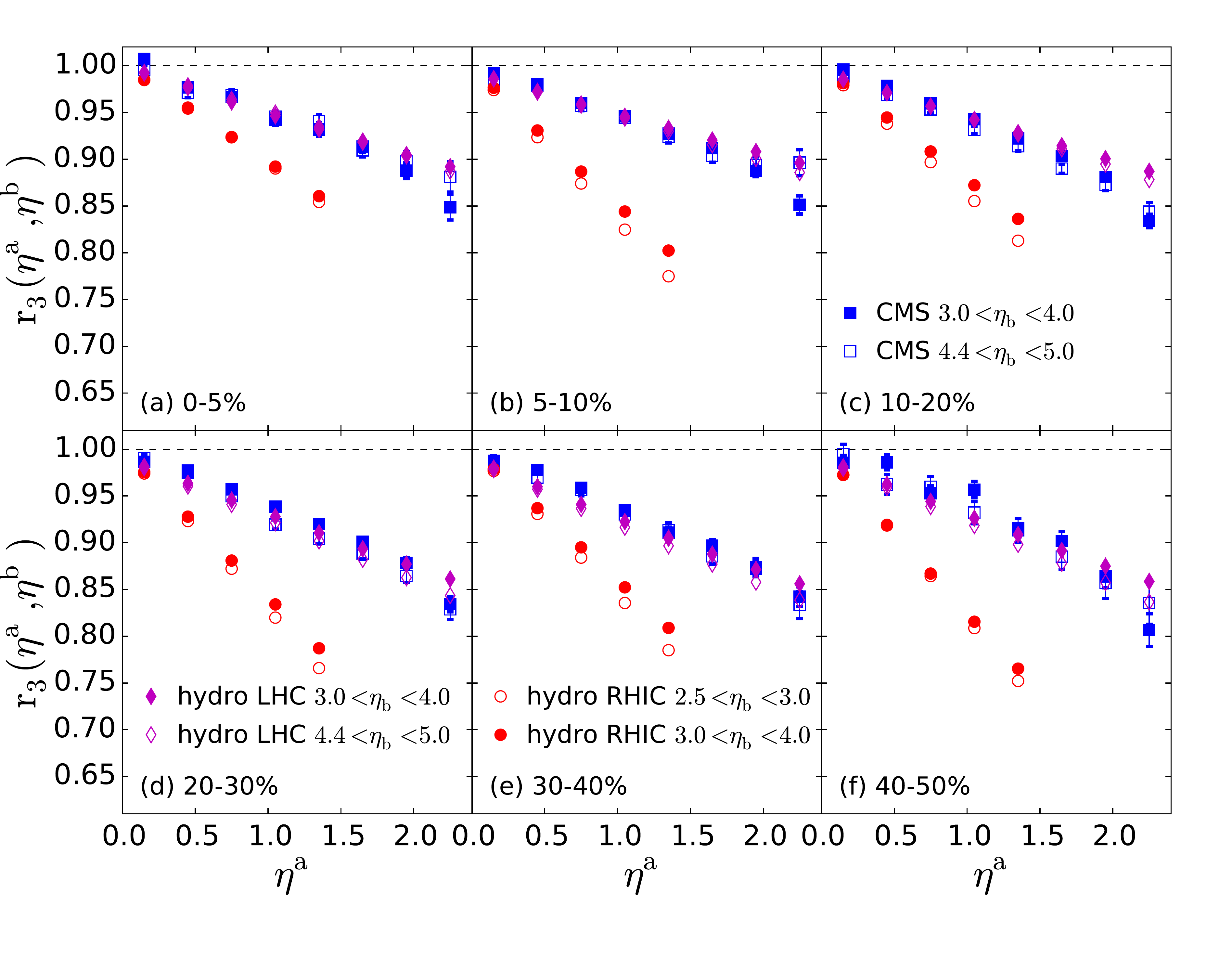} 
\protect\caption{(Color online) The same as Fig.~\ref{fig:r2} except for factorization ratio $r_{3}$.
\label{fig:r3}}
\end{figure}
\end{widetext}

It is interesting that the decorrelation in the factorization ratio as a function of $\eta^{a}$
at both RHIC and LHC energy have a linear behavior, which can
be parameterized with,
\begin{equation}
r_{n}(\eta^{a},\eta^{b})\approx e^{-2F_{n}^{\eta}\eta^{a}}\label{eq:twist_angle},
\end{equation}
where the factor $F_{n}^{\eta}$ can be considered as a measure of the factorization 
breakdown.  The decorrelation in the longitudinal direction can be caused by a systematic twist
and additional random fluctuations.  The twist of event planes originates from the forward-backward 
asymmetry for the transverse distribution of participant projectile and target nucleons. It is 
not expected to have a strong dependence on the beam energy.  The stronger decorrelation
we observe in Figs.~\ref{fig:r2} and \ref{fig:r3} at RHIC as compared to that at LHC is
caused mainly by larger fluctuations due to smaller number of initial partons. In the AMPT model which
uses HIJING for initial semi-hard jet production and soft string excitation, the smaller number of
initial partons at RHIC energy relative to LHC is due to smaller number of mini-jets and shorter length
of soft strings.  For lower beam energies at RHIC, fluctuations of the string length in the initial state
are the main mechanism for decorrelation in the longitudinal direction. This also explains the strong
decorrelation observed in the most central collisions. The experimental observation of such a stronger 
decorrelation in pseudo-rapidity at RHIC will provide another confirmation about the string picture of initial 
parton production. This picture captures most of the longitudinal fluctuations and correlations in coordinate space,
which are converted into the longitudinal decorrelation and long range correlation of final charged 
hadrons in momentum space.

\section{Longitudinal decorrelation in the initial state \label{sec:initial_state_decorr}}

To investigate further the origin of the final-state decorrelation
of event planes, we investigate fluctuations of the initial-state geometry
in terms of spatial eccentricity vectors at different space-time rapidity $\eta_{s}$,
\begin{equation}
\vec{\epsilon}_{n}(\eta_{s})=\epsilon_{n}\exp(i\Psi_{n})=\frac{\int d\mathbf{r_{\perp}}^{2}\varepsilon(r,\phi,\eta_{s})e^{in\phi}r^{n}}{\int d\mathbf{r_{\perp}}^{2}\varepsilon(r,\phi,\eta_{s})r^{n}}  . 
\label{eq:epsilon}
\end{equation}

In order to get a better linear correspondence between initial state eccentricity $\epsilon_n(\eta_s)$ and final-state
momentum anisotropy $v_n(\eta)$,
the distribution of local energy density $\varepsilon(r,\phi,\eta_{s})$ in each spatial rapidity window is recentered
according to the center of mass in this rapidity window. However, we should keep in mind that two adjacent fluid cells
in rapidity direction do interact with each other in (3+1)D hydrodynamics, which introduces non-linear correlation
between $\epsilon_n(\eta_s)$ and $v_n(\eta)$.
This is an intrinsic feature in (3+1)D hydrodynamics with longitudinal fluctuations.
One will indeed observe this feature in the comparison between longitudinal decorrelation of
initial state $\epsilon_3$ and final state $v_3$.

The decorrelation of initial eccentricities along spatial rapidity is defined analogously
to Eq.~(\ref{eq:r_n_final}) as,
\begin{equation}
r_{n}(\eta_{s}^{a},\eta_{s}^{b})=\frac{\left\langle \vec{\epsilon}_{n}(-\eta_{s}^{a})\cdot \vec{\epsilon}_{n}^{*}(\eta_{s}^{b})\right\rangle }{\left\langle \vec{\epsilon}_{n}(\eta_{s}^{a})\vec{\epsilon}_{n}^{*}(\eta_{s}^{b})\right\rangle } .
\label{eq:repsilon}
\end{equation}

\begin{figure}[!tph]
\includegraphics[width=0.5\textwidth]{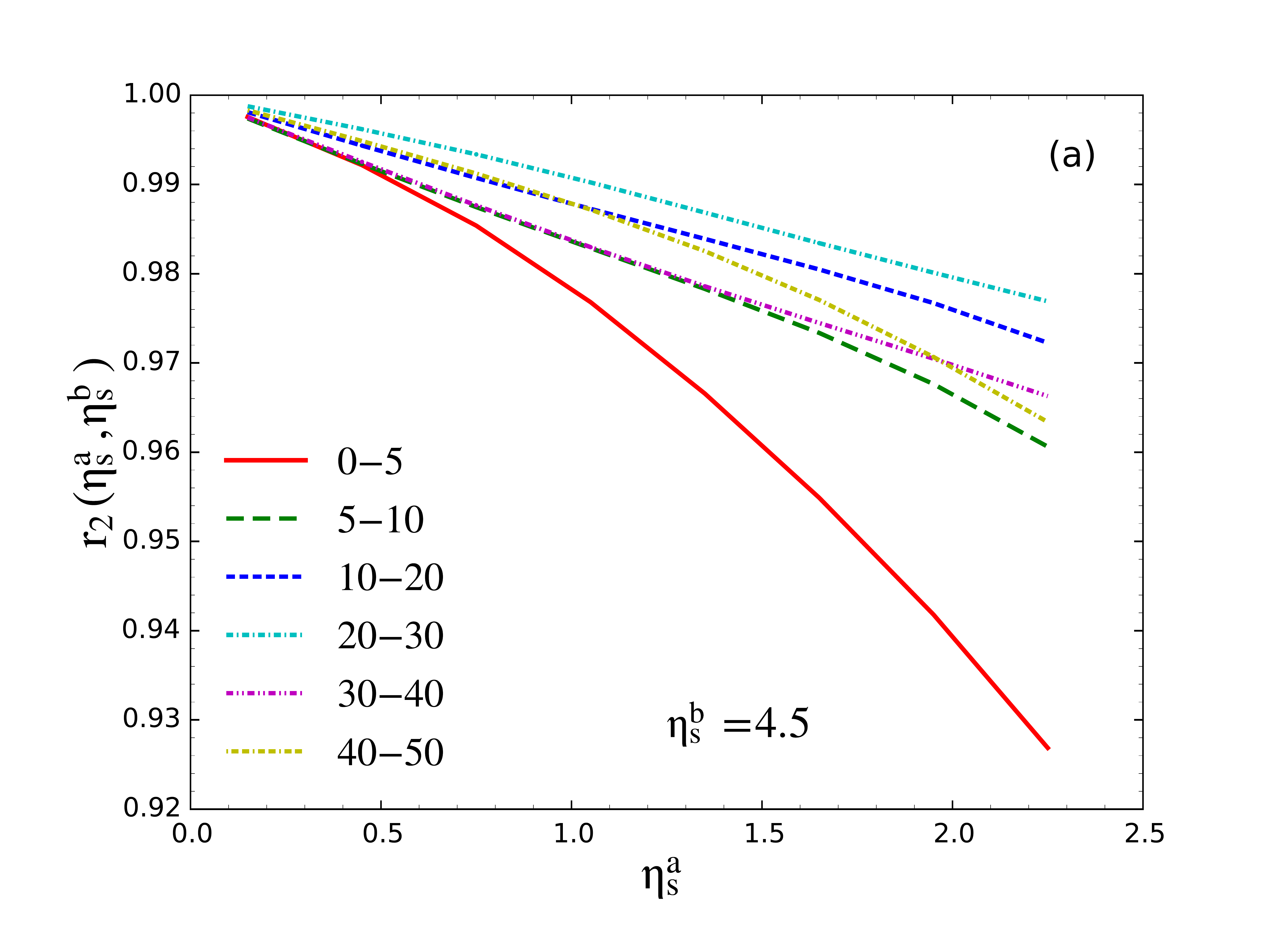} 
\vspace{-0.3in}

\includegraphics[width=0.5\textwidth]{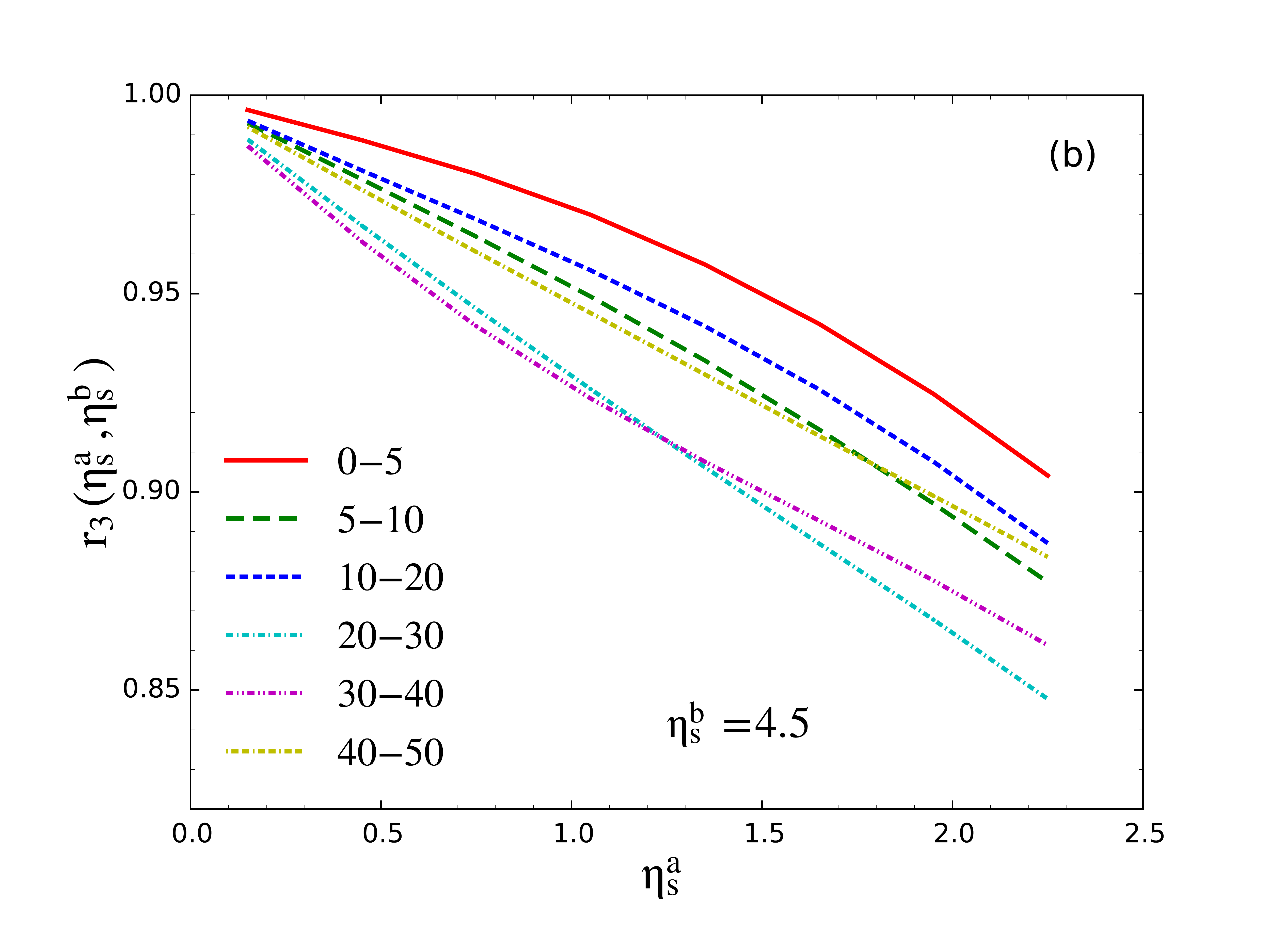} 

\protect\caption{(Color online) (a) The factorization ratio $r_{2}(\eta_{s}^{a},\eta_{s}^{b})$
 and (b) $r_{3}(\eta_{s}^{a},\eta_{s}^{b})$ for initial spatial eccentricity as a function of $\eta_{s}^{a}$
for $\eta_{s}^{b}=4.5$ in Pb+Pb collisions at $\sqrt{s_{NN}}=2.76$
TeV from AMPT model  in $6$ different centralities.
 \label{fig:r2_etas}}
\end{figure}

\begin{figure}[!tph]
\includegraphics[width=0.49\textwidth]{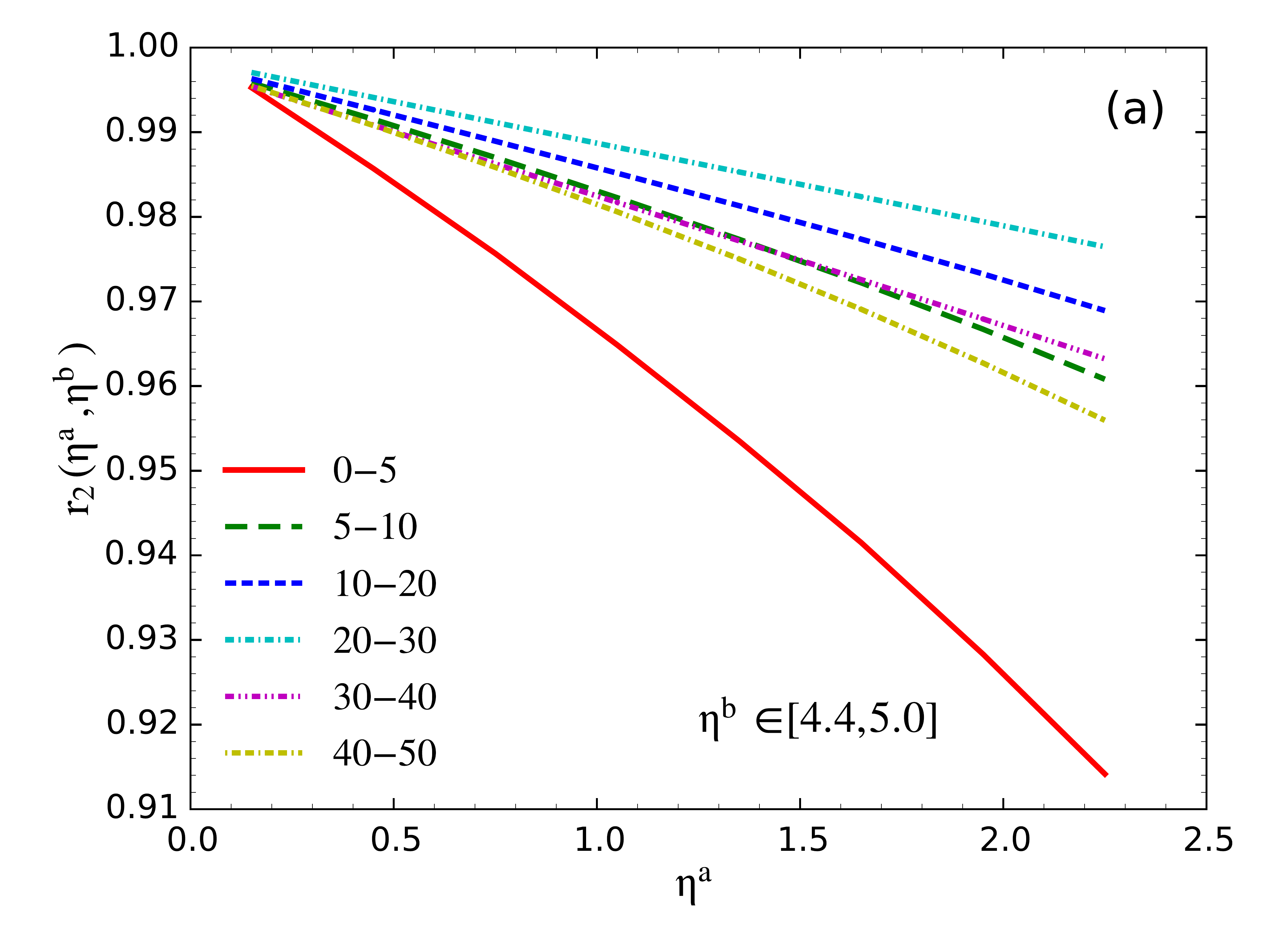}
\vspace{-0.3in}

\includegraphics[width=0.49\textwidth]{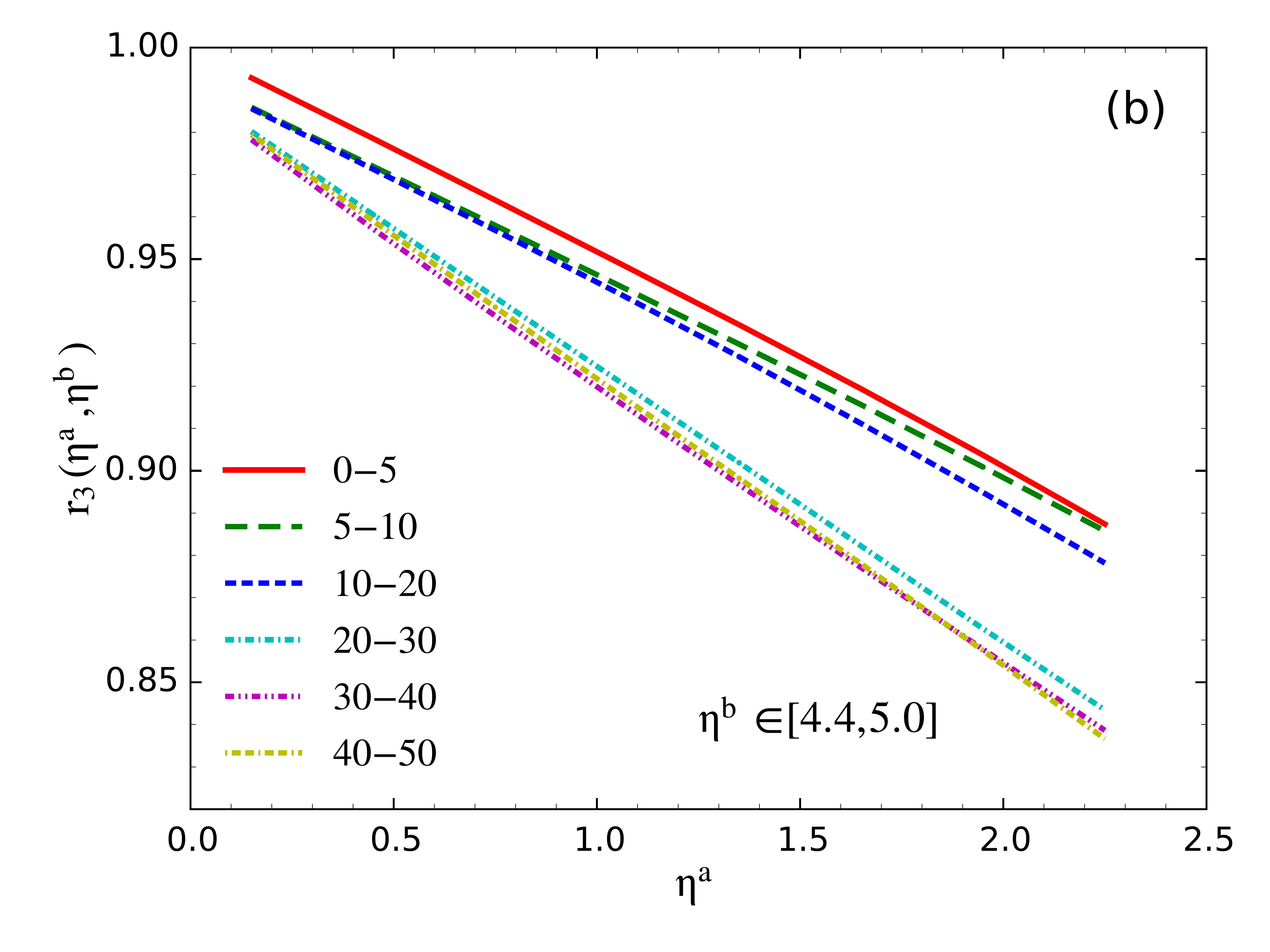} 
\protect\caption{(Color online) (a) The factorization ratio $r_{2}(\eta^{a},\eta^{b})$ 
and (b) $r_{3}(\eta^{a},\eta^{b})$ for final charged hadrons from event-by-event hydrodynamics as a function of the 
pseudo rapidity $\eta^{a}$ for $\eta^{b}\in[4.4,5.0]$ in Pb+Pb collisions at $\sqrt{s_{NN}}=2.76$ TeV
 for $6$ centrality classes.
\label{fig:rn_eta}}
\end{figure}

Shown in Fig.~\ref{fig:r2_etas}(a), the factorization ratio $r_{2}(\eta_{s}^{a},\eta_{s}^{b})$ 
in coordinate space displays the same centrality dependence as the 
decorrelation of final-state hadrons shown in Fig.~\ref{fig:rn_eta} for similar centralities of
Pb+Pb collisions at  $\sqrt{s_{NN}}=2.76$ TeV. For most central collisions, the decorrelation 
increases faster as $\eta_{s}^{a}$ increases. For other centralities, $r_{2}$ first increases from
central to semi-peripheral and reaches a maximum at $20-30\%$, then
it decreases for more peripheral collisions. The magnitude of the
longitudinal decorrelation $r_{2}$ in coordinate space is very close
to decorrelation of final charged hadrons. 
%The centrality dependence is similar to the centrality dependence of $p_{T}$ integrated elliptic flow.

The factorization ratio for the third-order initial eccentricity  $r_{3}(\eta_{s}^{a},\eta_{s}^{b})$ 
as shown in Fig.~\ref{fig:r2_etas} (b), on the other hand, decreases from the most central 
to peripheral collisions. The same centrality dependence is also observed in the factorization ratio of
the anisotropic flow in the final state as shown in Fig.~\ref{fig:rn_eta} (b). However, from 
mid-central to peripheral collisions there seems no particular
order in the centrality dependence of the factorization ratio $r_3$ for the initial eccentricity. 
%The hydrodynamic expansion is more important
%here to convert the correlation of initial eccentricity in coordinate space to a correlation
%of anisotropic flow in the final state, which reduces
%the centrality dependence through out the hydrodynamic evolution. 

\begin{figure}[!tph]
\includegraphics[width=0.5\textwidth]{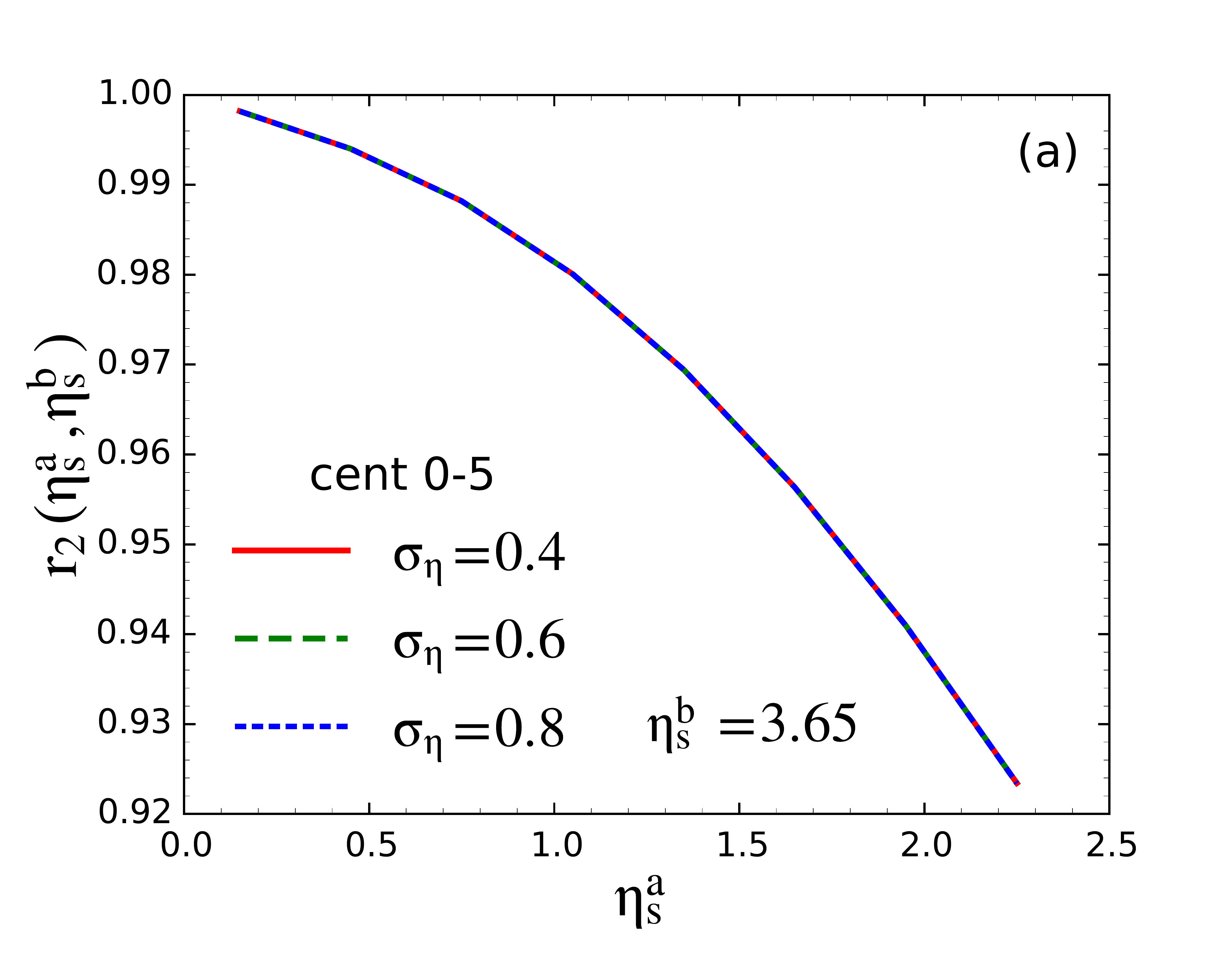}\\
\vspace{-0.40in}

\includegraphics[width=0.5\textwidth]{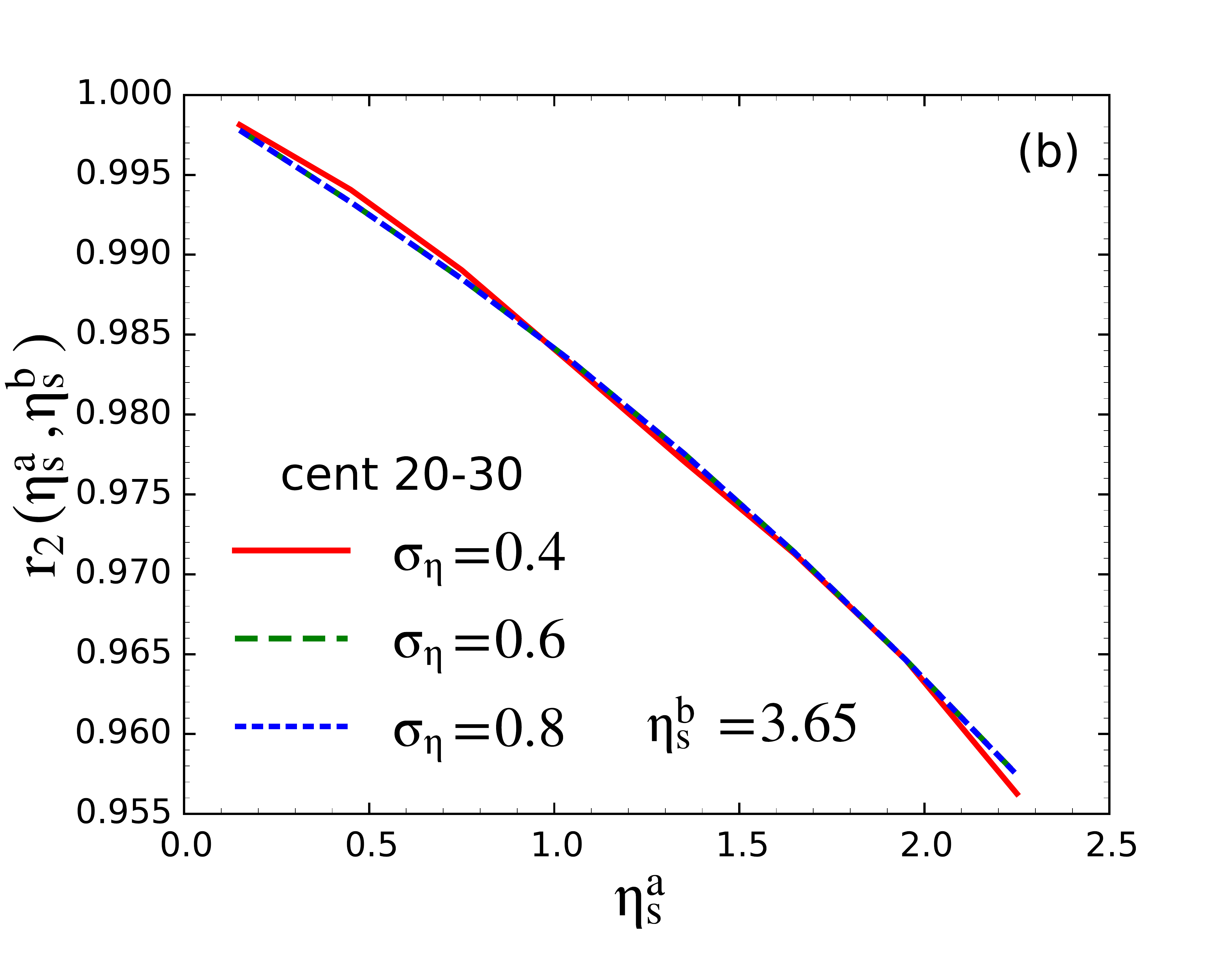} 
\protect\caption{(Color online) The decorrelation function
$r_{2}(\eta_{s}^{a},\eta_{s}^{b})$ along spatial rapidity for different
values of the width $\sigma_{\eta_s}$ of the Gaussian smearing in the initial-state 
of Pb+Pb collisions at $\sqrt{s_{NN}}=2.76$ TeV from the AMPT model in the top
(a) 0-5\% and 20-30\% centrality.}

 \label{fig:smearing_width}

\end{figure}

The definition of the factorization ratio is designed to remove
short range correlations by correlating particles with
large pseudo rapidity gaps. Therefore, the decorrelation of anisotropic
flows along the longitudinal direction should not depend on the width of the 
Gaussian smearing in spatial rapidity in initial conditions from the AMPT model.
Shown in Fig.~\ref{fig:smearing_width} are factorization ratios for the initial
eccentricity $r_{2}(\eta_{s}^{a},\eta_{s}^{b})$ as a function of $\eta_{s}^a$ in
two centralities of Pb+Pb collisions at $\sqrt{s_{NN}}=2.76$ TeV with
different values of the width $\sigma_{\eta}=0.4,\ 0.6,\ 0.8$ in 
the Gaussian smearing. Indeed, the factorization ratio shows no
dependence on the smearing width.

\section{Twist or random fluctuations \label{sec:twist_or_fluctuation}}

There are two possible mechanisms for the decorrelation along the longitudinal
direction. One is the twist of event plane angles and the other are
random fluctuations along pseudo rapidity. In event-by-event (3+1)D
hydrodynamics, without fluctuations due to finite number of particles in the final state,
we can calculate event plane angles at different pseudo-rapidities
$\Psi_{n}(\eta)$ for each event, and study the fluctuation or
twist structure for $\Psi_{n}(\eta)$.

\begin{figure}[!tph]

\includegraphics[width=0.5\textwidth]{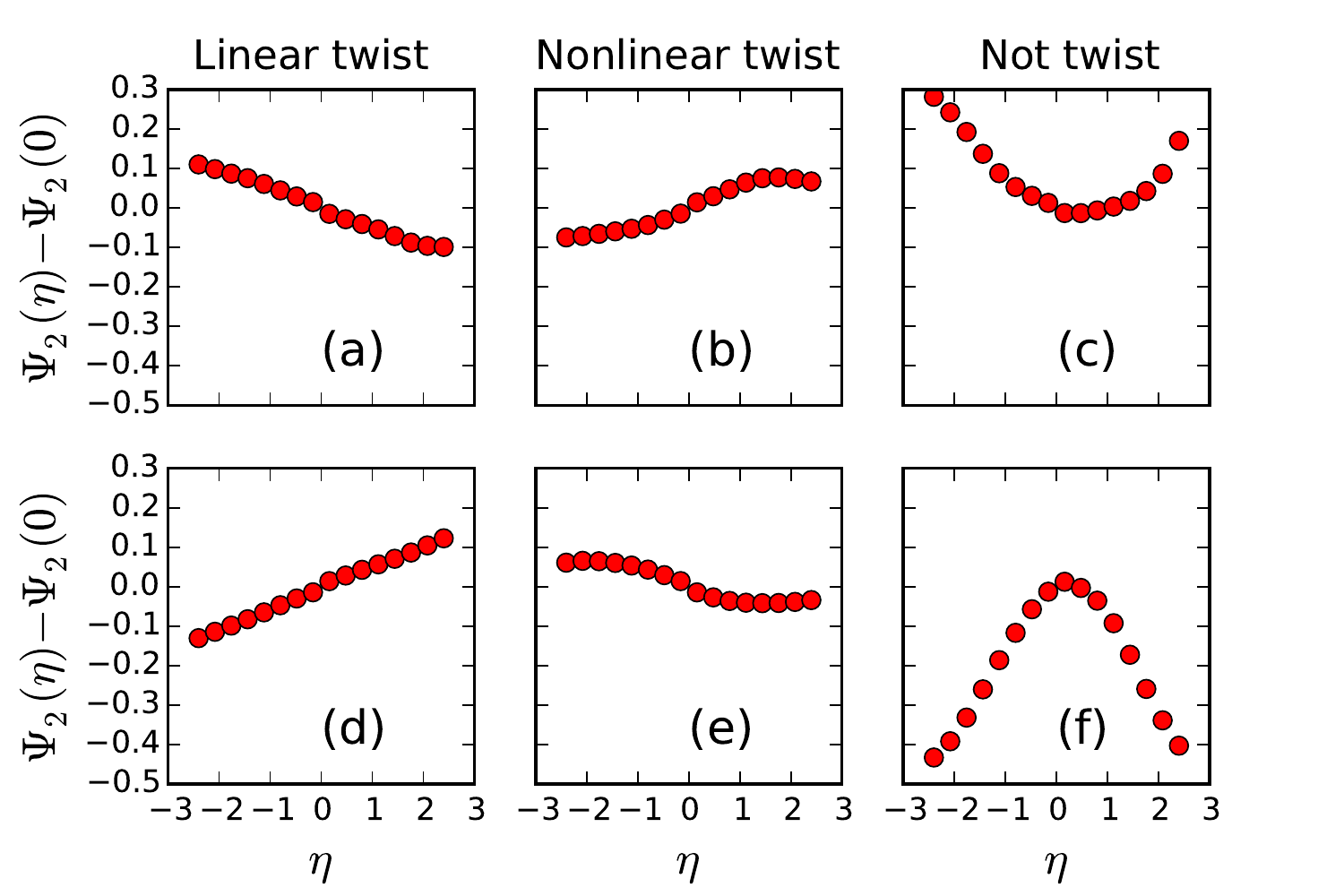}
\protect\caption{(Color online) The $2$nd order event plane angles $\Psi_{2}(\eta)$
as a function of pseudo-rapidity $\eta$ for $6$ typically selected
events from event-by-event hydrodynamic simulations for $0-5\%$ Pb+Pb
collisions at $\sqrt{s_{NN}}=2.76$ TeV. \label{fig:psin_eta}}

\end{figure}

Shown in Fig.~\ref{fig:psin_eta} are event plane angles for 6 events in the top 0-5\% 
centrality class from our (3+1)D hydrodynamic calculations as a function of $\eta$
in Pb+Pb collisions at $\sqrt{s_{NN}}=2.76$ TeV. Some events have a clear twist structure, i.e.,  
event plane angles $\Psi_{n}$ monotonically increase or decrease along $\eta$ direction. 
Some events do not have an apparent twist structure. Their event plane angles
do not vary with $\eta$ monotonically.

\begin{figure}[!tph]

\includegraphics[width=0.4\textwidth]{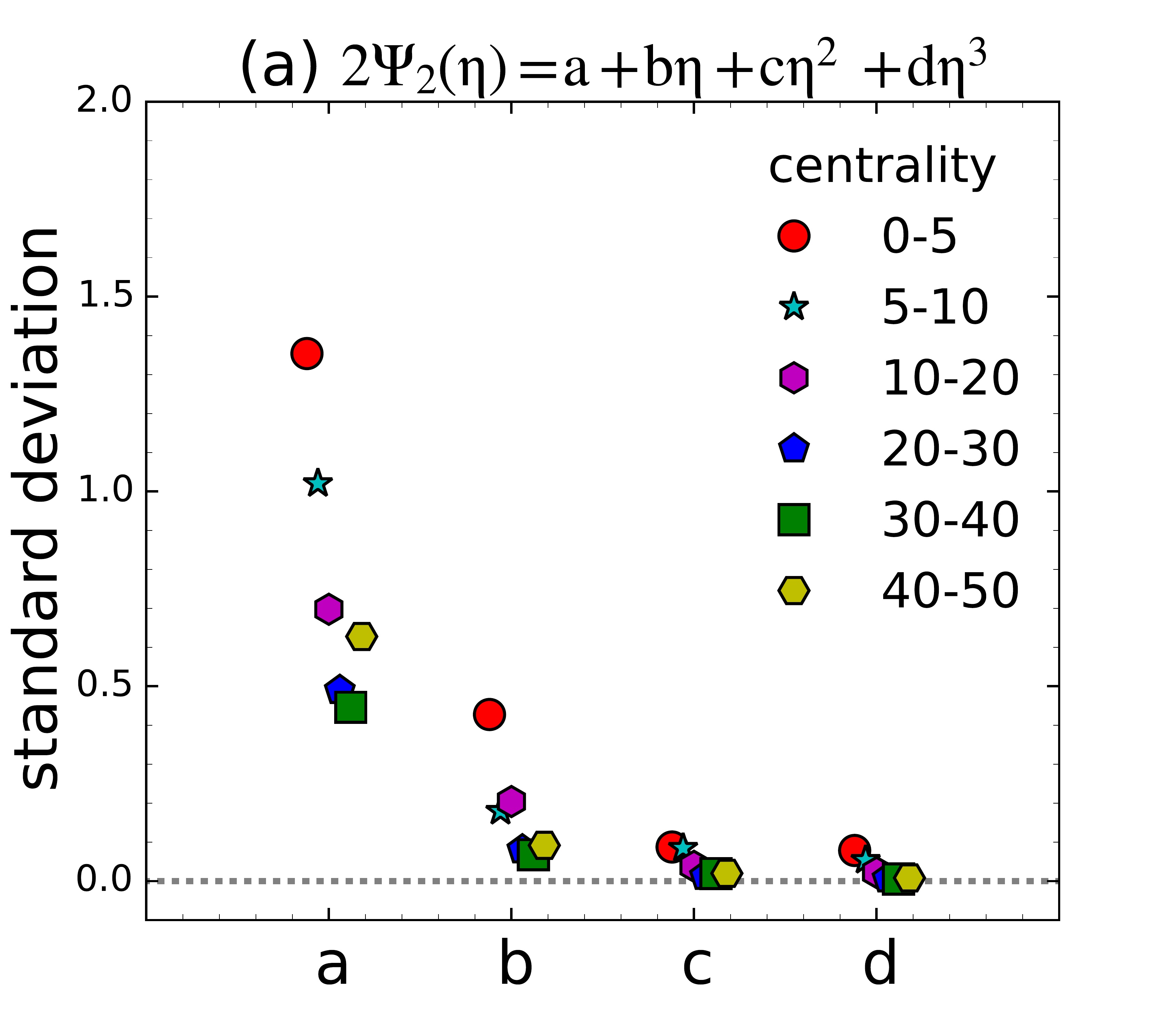}
\includegraphics[width=0.4\textwidth]{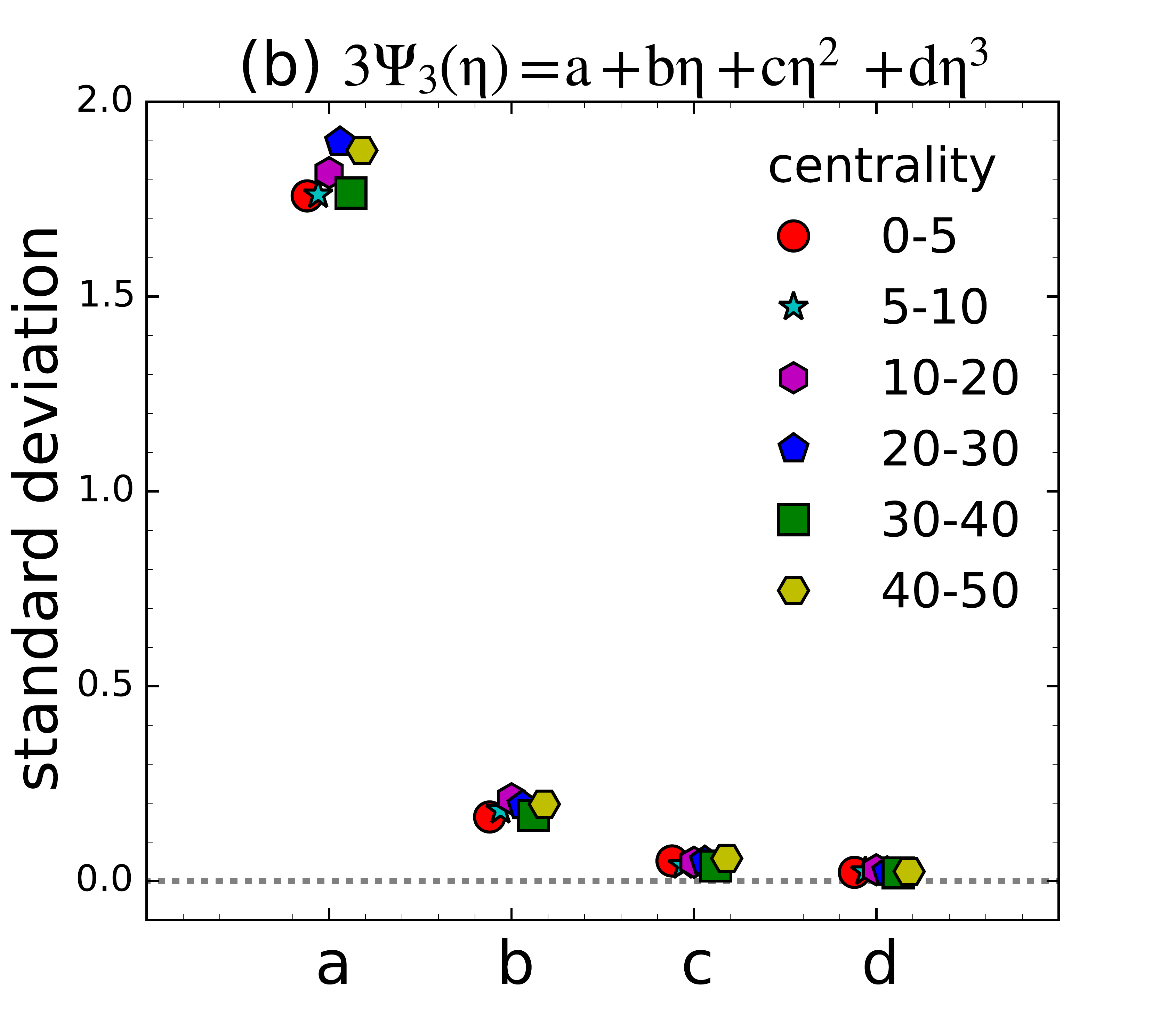}

\protect\caption{(Color Online) Standard deviation of fitting polynomial coefficients
for $\Psi_{2}(\eta)$ and $\psi_{3}(\eta)$.}
\label{fig:std}
\end{figure}

Typically there are events with a  linear twist, nonlinear twist and events without twist as illustrated 
in Fig.~\ref{fig:psin_eta}. Probabilities for these $3$ different types of events are studied by fitting 
the event plane angle $\Psi_{n}(\eta)$ with a polynomial
function $\Psi_{n}(\eta)=a+b\eta+c\eta^{2}+d\eta^{3}$, where non-zero
$b$ denotes a linear twist of event plane angles. Non-zero polynomial
coefficients $c$ and $d$ denote non-linear twist or pure fluctuations
of event plane angles. Since the event averaged values of $b$, $c$ and
$d$ are equal to $0$, we can use the standard deviations $\sigma(x)=\langle(x-<x>)^{2}\rangle$ 
of these polynomial coefficients as a measure of fluctuation and decorrelation,
as shown in Fig.~\ref{fig:std} for 6 centrality classes of Pb+Pb collisions
at  $\sqrt{s_{NN}}=2.76$ TeV. 

For the most central $0-5\%$ collisions, the standard deviations
of $a,b,c$ and $d$ in the 2nd order event plane angles are much bigger than
in semi-central collisions. Large values of  $\sigma(a)$ indicate
strong event-by-event fluctuations of the mean values of the event plane 
angle. One can see from Fig.~\ref{fig:std} (a)  that  the $2$nd order
event plane angles fluctuate more  in central collisions than in semi-central 
and peripheral collisions. A large value of $\sigma(b)$ indicates that a strong 
decorrelation along the longitudinal direction comes from a linear twist of 
the event plane angles. The twist of the 2nd order event plane angles
also have a strong centrality dependence as seen in Fig.~\ref{fig:std} (a).
On the other hand, standard deviations for coefficients of a polynomial fit to the $3rd$ event 
plane angles do not have any significant centrality dependence as shown in Fig.~\ref{fig:std}  (b).

Finite values of $\sigma(c)$ and $\sigma(d)$ indicates finite probabilities  for the non-linear
twist and pure fluctuations of event plane angles  which are also bigger
in the most central collisions.  Over all,  small standard deviations for $c$ and $d$ from
both $\Psi_{2}(\eta)$ in semi-central collisions and $\Psi_{3}(\eta)$
in all centralities explain the linear behavior of the longitudinal
decorrelation in these centralities.  

%In the central collisions, the measured $r_{3}(\Delta\eta)$ is stronger than $r_{2}(\Delta\eta)$ in CMS,
%but they are quite similar in hydrodynamics. However, the twist of event planes for $\Psi_{3}$
%is smaller than $\Psi_{2}$ in hydro, which indicates that fluctuations of $v_{3}$ is bigger than $v_{2}$
%along the longitudinal direction.

Although hydrodynamic results underestimated $r_{2}$ given by CMS experimental 
data in most central collisions, it has a similar feature as in CMS measurements that the decorrelation
$r_{2}$ is strongest in most central collisions. The discrepancy between the (3+1)D ideal hydrodynamic
results and the CMS experimental data on decorrelation of the second order anisotropic flow 
in the most central collisions might be reduced by the introduction of viscosity
in the hydrodynamic calculation.  Since the shear viscosity reduces
the expansion rate along $\eta_{s}$ direction, it is possible that 
the decorrelation structure along the longitudinal direction for
central collisions will also change.

The centrality dependence of the twist effect
on the other hand agrees with predictions from the Glauber model \cite{Jia:2014ysa},
where the twist is strongest in most central collisions.

\section {Summary and discussions \label{sec:discussion}}

The decorrelation of $2$nd and $3$rd order anisotropic flow along
the pseudo rapidity direction is investigated in event-by-event (3+1)D
ideal hydrodynamics with fluctuating initial conditions from the AMPT model
for Pb+Pb collisions at LHC and Au+Au collisions
at RHIC. With transverse and longitudinal fluctuations in AMPT model
originating from pQCD+MC Glauber and string model, our results
agree with CMS measurements for most of the available centralities.
This suggests that the string model in HIJING model captures most of 
features of the longitudinal fluctuations along pseudo rapidity. Predictions
for Au+Au collisions at RHIC show stronger longitudinal decorrelations,
indicating larger fluctuations at lower energies. Further
detailed study show that the longitudinal decorrelation in momentum
space comes from initial state decorrelation in space. 

In order to explain the non-linear behavior of $r_{2}(\eta^{a},\eta^{b})$
in the most central collisions, event plane angles $\Psi_{n}(\eta)$
for particles at different pseudo rapidities are calculated from event-by-event
hydrodynamics for several characteristic  events in the most central collisions. 
Some of these events show non-linear twist and pure fluctuations of event plane angles
which provide the explanation for the non-linear behavior of $r_{2}$ in most central collisions.
By fitting $\Psi_{n}(\eta)$ with a polynomial function, contributions from
linear twist, non-linear twist and pure fluctuations are quantified
by the standard deviations of polynomial coefficients. Larger standard
deviations of non-linear polynomial coefficients in most central collisions
make it clear that the non-linear twist and pure fluctuations are
much bigger for $r_{2}$ in $0-5\%$ collisions. While very small
non-linear polynomial coefficients in all centralities for $\Psi_{3}(\eta)$
proves that the decorrelation of the $3$rd order anisotropic flow comes
mainly from linear twist of event plane angles.

The decorrelation of event plane angles and anisotropic flows along
the longitudinal direction discussed in this paper have many consequences in the study
of collective phenomena in high-energy collisions.\\

\noindent {\it  $v_{n}$ measurements}\\

Determination of event planes is crucial for $v_{n}$ calculations. Most of current analysis
methods use particles at forward or backward rapidity to determine event planes and calculate 
anisotropic flows for particles at mid rapidity. This method eliminates contributions to the anisotropic 
flows from short range correlations due to non-flow effects such as jets or fluid expansion of hot spots. 
These methods, however, assume rapidity-independent event planes and become problematic when 
event planes in different rapidity windows randomly fluctuate or are twisted.

The difference between event planes at $\eta_{a}$ and $-\eta_{a}$ has been attributed to the effect of 
finite multiplicities. This was assumed to be corrected by dividing the anisotropic flow
 $\left\langle \cos\left(n(\phi-\Psi_{n}^{A})\right)\right\rangle $ by a resolution factor 
 \begin{equation}
 R=\sqrt{\frac{\left\langle \cos\left(n\left(\Psi_{n}^{A}-\Psi_{n}^{B}\right)\right)\right\rangle \left\langle \cos\left(n\left(\Psi_{n}^{A}-\Psi_{n}^{C}\right)\right)\right\rangle }{\left\langle \cos\left(n\left(\Psi_{n}^{B}-\Psi_{n}^{C}\right)\right)\right\rangle }},
 \end{equation}
where $\Psi_{n}^{A}$ and $\Psi_{n}^{B}$ are event plane angles determined
from particles at forward and backward rapidity windows and $\Psi_{C}$
is the event plane angle at mid rapidity \cite{Jia:2014ysa}. If event plane angles
are linearly twisted along the pseudo rapidity direction, the decorrelation
between $\Psi_{n}^{A}$ and $\Psi_{n}^{C}$ equals to the decorrelation
between $\Psi_{n}^{B}$ and $\Psi_{n}^{C}$, the resolution would
become
\begin{eqnarray}
R&=&R_{0}\sqrt{\left\langle \cos\left(n\left(\Psi_{n}^{+}-\Psi_{n}^{-}\right)\right)\right\rangle } \nonumber \\
&\approx & R_{0}\left\langle \cos\left(n\left(\Psi_{n}^{+}-\Psi_{n}^{0}\right)\right)\right\rangle ,
\end{eqnarray}
where $R_{0}$ comes from the effect of a finite multiplicity
while $\left\langle \cos\left(n\left(\Psi_{n}^{+}-\Psi_{n}^{0}\right)\right)\right\rangle $
stands for the decorrelation of the true event plane $\Psi_{n}^{+}$
at forward rapidity and $\Psi_{n}^{0}$ at mid rapidity. Approximately,
currently used resolution factors give $v_{n}=\left\langle \cos\left(n(\phi-\Psi_{n}^{0}\right)\right\rangle $,
if the structure of longitudinal fluctuations is a global twist \cite{WeiLi:2015int}.
In this sense the event plane method is better than 2-particle correlation method in which a
large pseudo rapidity gap is used to remove non-flow effects.
The method using multiple particle cumulants without requiring a large pseudo rapidity gap
is better for flow measurements.
%especially for low beam energies and small systems.

If event planes in the final state are linearly twisted, which can be parameterized as $\Psi_{n}(\eta)=b\eta$, and
the magnitude of $v_{n}$ is the same for each rapidity slice in a rapidity window $[-1,1]$,
using the mean value of the event plane angle $\Psi_{n}(\eta)=0$ for all particles
in this rapidity window will reduce $v_{n}$ to $v_{n}\sin(nb)/(nb)$.
The factor $\sin(nb)/(nb)$ depends on the slope of the twist in $\Psi_{n}(\eta)$ and the order of 
harmonic flows.\\

\noindent {\it Effects of longitudinal expansion.}\\

In a previous study \cite{Pang:2012he}, by changing the width of the rapidity
window for determining event planes in (3+1)D hydrodynamics calculations, only a $6\%$
difference has been observed coming from the decorrelation of event planes, the other
$10-15\%$ suppression of the elliptic flow originates from the coupling
between transverse and longitudinal expansion. The additional gradient
in the longitudinal direction between two adjacent  spatial rapidity windows
which have different spatial eccentricities and orientation angles
can lead to an overall reduction in eccentricity and anisotropy flows of final particles.
Notice that the measured decorrelation of anisotropic flows along $\eta$
is in momentum space. How such decorrelation is influenced by hydrodynamic expansion 
in longitudinal direction with fluctuating initial conditions and the effect of viscosity 
are interesting topics for further exploration.\\

\noindent {\it Long range di-hadron correlation in $\Delta\phi$}\\

Another observable that will be affected by the decorrelation of anisotropic
flows is the di-hadron correlation as a function of azimuthal angle
difference $\Delta\phi$ and pseudo rapidity difference $\Delta\eta$.
With a linear twist in event plane angles $\Psi_{n}(\eta)$, the di-hadron
correlation function $C_{12}(\Delta\eta,\Delta\phi)$ will have an
intrinsic structure for the near-side and away-side ridge. In each
single event, the ridge will be shifted from $\Delta\phi=0$ to $\Delta\phi=2b\Delta\eta$
on the near side and from $\Delta\phi=\pi$ to $\Delta\phi=\pi+2b\Delta\eta$
on the away side, where $b$ is the slope in the linear twist in $\Psi_{n}(\eta)=b\eta$.
Since the slope $b$ can be positive or negative, the event averaged
di-hadron correlation will be broadened along $\Delta\phi$ with a big
pseudo rapidity gap \cite{Jia:2014ysa}.

\begin{acknowledgments}
We thank J. Jia, W. Li, P. Bozek, P. Huovenien and M. Gyulassy for helpful
discussions. LGP and HP acknowledge funding
of a Helmholtz Young Investigator Group VH-NG-822 from the Helmholtz
Association and GSI. This work was supported in part by the Helmholtz International
Center for the Facility for Antiproton and Ion Research (HIC for FAIR)
within the framework of the Landes-Offensive zur Entwicklung Wissenschaftlich-Oekonomischer
Exzellenz (LOEWE) program launched by the State of Hesse, 
by the Natural Science Foundation of China under Grants No. 11221504 and No. 11375072, by the Chinese Ministry of Science and Technology under Grant No. 2014DFG02050, and by the Director, Office of Energy Research, Office of High Energy and Nuclear Physics, Division of Nuclear Physics, of the U.S. Department of Energy under Contract No. DE- AC02-05CH11231.
V.R. is supported by the Alexander von Humboldt foundation, Germany.
Computational resources have been provided by the Center for Scientific Computing
(CSC) at the Goethe-University of Frankfurt.
\end{acknowledgments}

\bibliographystyle{unsrt}
\bibliography{ref}

\end{document}